\documentclass[12pt]{article}

\usepackage{graphicx}

\begin{document}
\title{\bfseries
Quantum and classical advantage distillation are not equivalent}

\author{%
Dagomir \textsc{Kaszlikowski}$^1$, %
\textsc{Lim} Jenn Yang$^1$, %
\\
\textsc{Kwek} Leong Chuang$^{2,1}$, %
Berthold-Georg \textsc{Englert}$^1$ \\[1ex]
\itshape\normalsize%
$^1$Department of Physics, National University of Singapore,\\
\itshape\normalsize
Singapore 117542, Singapore\\
\itshape\normalsize
$^2$National Institute of Education, %
Nanyang Technological University,\\
\itshape\normalsize Singapore 639798, Singapore}

\date{\normalsize(24 October 2003)}

\maketitle

\begin{abstract}
We report that, for the generation of a secure cryptographic key from
correlations established through a noisy quantum channel, the quantum and
classical advantage distillation procedures are not equivalent, when coherent
eavesdropping attacks are duly taken into account. 
The quantum procedure can tolerate markedly more noise in the channel than
the classical procedure.
\end{abstract}

One of the important problems in quantum cryptography is whether
classical methods of key distillation are equivalent to quantum methods.
This problem was recently investigated for a certain class of 
protocols in Refs.\ \cite{US,THEM,YC}. 
It was concluded there that the quantum entanglement distillation \cite{HOR1} 
tolerates exactly the same amount of errors in a raw cryptographic key as 
classical advantage distillation \cite{MAURER}. 
The analysis was carried out under the assumption that Eve performs the most
general incoherent attack.  

In this short note \cite{footnote} we report that this equivalence is only
apparent, in fact classical advantage distillation is not as powerful as
quantum entanglement distillation.
This is so because Eve has the option of coherent attacks which give her more
information than what she can get from incoherent attacks.
Roughly speaking, Eve exploits the classical information she gains from the
public communication exchanged between Alice and Bob as part of the advantage
distillation protocol.

Alice sends \emph{qunits} to Bob, that is $n$-dimensional quantum systems,
and the raw key sequence consists of the paired nit values for matching
measurement bases.
The security of the protocols of \cite{US,THEM,YC} for quantum cryptography
can be expressed solely in terms of the probability $\beta_0$ that Alice and
Bob get the same nit value if the bases match. 
(We are adopting the notation of Ref.\ \cite{US}.)
For an ideal noiseless quantum channel, one has $\beta_0=1$, and $\beta_0=1/n$
obtains for a channel that has nothing but unbiased noise.
Channels with some admixture of noise are characterized by $\beta_0$ values
between those extremes. 

The lesson of Refs. \cite{Bruss+1:02,THEM,US} is that classical advantage
distillation can be performed successfully if
\begin{equation}
  \label{eq:old}
  \beta_0>\frac{2}{2+(n-1)}\equiv\beta_0^\mathrm{(inc)}
\quad\mbox{(old threshold)}
\end{equation}
which is also the threshold for quantum entanglement distillation \cite{HOR1}.
This result is valid \emph{provided that} incoherent eavesdropping attacks 
supply as much information as possible.
This assumption is not implausible, inasmuch as Gisin and Cirac \cite{G+C},
and Wang \cite{Wang} argued that coherent eavesdropping attacks cannot be more
powerful than incoherent ones.

\begin{figure}[!t]
\centering
\begin{picture}(380,270)(0,-5)
\put(20,10){\includegraphics{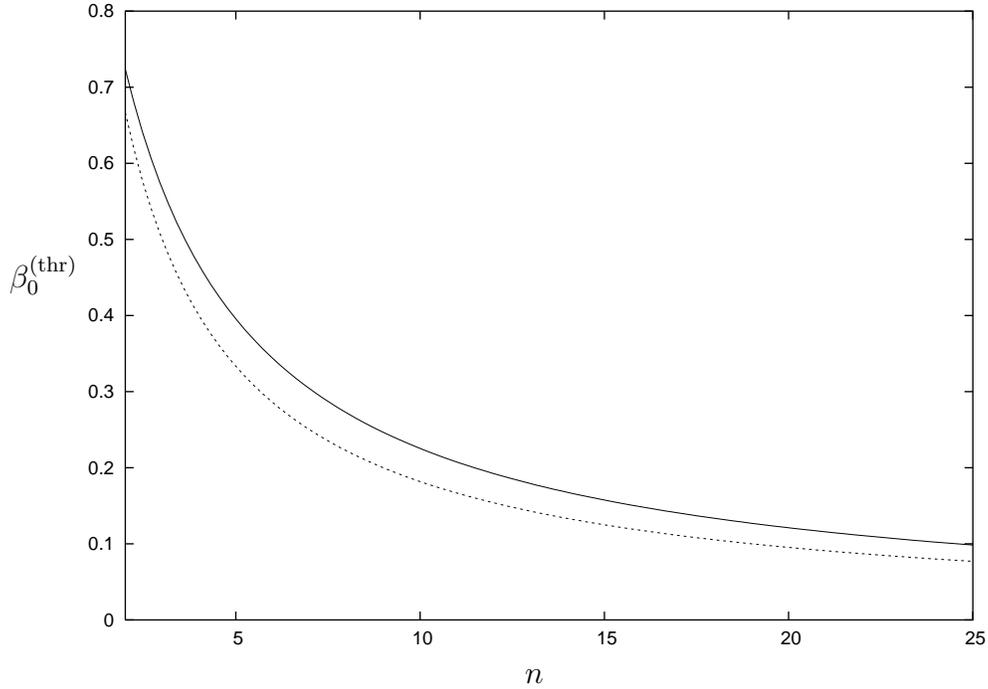}}
\put(200,0){$n$}
\put(5,150){$\beta_0^\mathrm{(thr)}$}
\end{picture}
\caption{\label{fig}%
Threshold values $\beta_0^\mathrm{(thr)}$ for incoherent attacks 
($\beta_0^\mathrm{(thr)}=\beta_0^\mathrm{(inc)}$, dashed
line) and coherent attacks ($\beta_0^\mathrm{(thr)}=\beta_0^\mathrm{(coh)}$, 
solid line), for $2\leq n\leq25$.
} 
\end{figure}

In the context of advantage distillation, however, there is a self-suggesting
coherent attack that truly outperforms the best incoherent attack. 
Instead of measuring her ancillas one by one, Eve 
performs a collective measurement on a naturally chosen subset.
We analyzed this scenario in full and find that the threshold of
(\ref{eq:old}) is too optimistic. 
Rather, the true threshold is given by
\begin{equation}
  \label{eq:new}
   \beta_0>\frac{2}{2+(3-\sqrt{5})(n-1)}\equiv\beta_0^\mathrm{(coh)}
\quad\mbox{(new threshold)}. 
\end{equation}
This is a substantially more stringent condition on $\beta_0$ than 
the one in (\ref{eq:old}) that is derived for incoherent attacks only; 
see Fig.~\ref{fig}.

We are indebted to Ajay Gopinathan, Christian Kurtsiefer, 
Yeong Cherng Liang and Christian Miniatura for sharing their insights with
us. 
This work was supported by A$^*$Star Grant No.~012-104-0040.

\renewcommand{\refname}   


\begin{thebibliography}{99}
\bibitem{footnote} 
A more detailed account is being prepared.
\end{thebibliography}
\end{document}